\documentstyle[epsfig]{elsart}
\begin{document}
\begin{frontmatter}
\title{ On the Formation of
Degenerate Heavy Neutrino Stars}
\author{Neven Bili\'c\thanksref{zagreb}},
\author{Robert J. Lindebaum\thanksref{rjladd}},
\author{Gary B. Tupper},
\author{Raoul D. Viollier\thanksref{cor}}
\address{Institute of Theoretical Physics and Astrophysics,
 Department of Physics, University of Cape Town,
 Private Bag, Rondebosch 7701, South Africa}
\thanks[zagreb]{Permanent address:
Rudjer Bo\v skovi\'{c} Institute,
P.O. Box 180, 10002 Zagreb, Croatia;
        Email: bilic@thphys.irb.hr}
\thanks[rjladd]{Permanent address:
School of Chemical and Physical Sciences,
 Uni\-ver\-si\-ty of Natal, Private Bag X01,
        Scottsville 3209, South Africa;
       Email: lindebaumr@nu.ac.za}
\thanks[cor]{Corresponding author; Email: viollier@physci.uct.ac.za}
\begin{abstract}
The dynamics of a self-gravitating cold Fermi gas is described
using the analogy with an interacting self-gravitating
Bose condensate having the same Thomas-Fermi limit.
The dissipationless formation of a heavy neutrino star
through gravitational collapse
and ejection of matter
is demonstrated numerically.
Such neutrino stars offer an alternative to
black holes for the supermassive
compact dark objects at the centers of galaxies.
\end{abstract}
\begin{keyword}
neutrino stars \sep boson stars \sep
self-gravitating systems \sep galactic centers
\PACS
 02.60.Cb \sep 95.30.Lz \sep 95.35.+d \sep 98.35.Jk
\end{keyword}
\end{frontmatter}

Supermassive neutrino stars, in which self-gravity is balanced by the
degeneracy pressure of the cold fermions, have been a subject of
speculation for more than three decades \cite{1}.  Originally, these
objects were proposed as models for dark matter in galactic halos and
clusters of galaxies, with neutrino masses in the $\sim$ eV
range. More recently, however, degenerate superstars composed of
weakly interacting fermions in the $\sim$ 10 keV range, have been
suggested as an alternative to the supermassive black holes that are
purported to exist at the centers of galaxies \cite{2,3,4,5,6}.
In fact, it has been shown \cite{4} that such degenerate fermion stars
could explain the whole range of supermassive compact dark objects
which have been observed so far, with masses ranging from $10^{6}$ to
$3 \times 10^{9}$$M_{\odot}$, merely assuming that a weakly
interacting quasi-stable fermion of mass $m_{f} \simeq$ 15 keV exists
in nature.

As an example, the most massive compact dark object ever
observed, is located at the center of M87, with a mass $M \simeq 3.2
\times 10^{9} M_{\odot}$ \cite{7}.
Interpreting this object as a relativistic
fermion star at the Op\-pen\-hei\-mer-Volkoff \cite{8} limit, yields a
fermion mass of $m_{f} \simeq 15$ keV and
 a fermion star radius of $R = 4.45 R_{\mathrm{S}}
\simeq 1.5$ light-days \cite{3,4}, where
$R_{\mathrm{S}}$ is the Schwarzschild
radius.
In this case there is little difference between the fermion
star and black hole scenarios,
 because the radius of the last stable orbit around a
Schwarzschild black hole is  $R=3R_{\mathrm{S}}$ anyway.

Extrapolating this down to the compact dark object at the center of
our galaxy \cite{9}, which,
having a
mass $M \simeq 2.6 \times 10^{6} M_{\odot}$,
is at the lower limit of the mass range
of the observed compact dark objects,
we obtain, using
the same fermion mass,
a radius
$R\simeq 20$ light-days $\simeq 5\times10^{4} R_{\mathrm{S}}$
\cite{2}.
As the potential inside such a nonrelativistic fermion star is shallow,
the spectrum of radiation emitted by accreting baryonic matter is cut
off for frequencies larger than $10^{13}$ Hz \cite{3,5}, as is
actually
observed in the spectrum of the
enigmatic radio source Sgr A$^{*}$ at the
galactic center \cite{10}.
A fermion star
with radius $R$
 \raisebox{-1.5mm}{$\stackrel{\textstyle <}{\sim}$}
20 light-days and mass
$M \simeq 2.6 \times 10^6 M_{\odot}$
is also consistent
\cite{6} with the observed motion of stars within a projected distance
of 6 to 30 light-days from Sgr A$^{*}$ \cite{9}.

Of course, it is well-known that 15 keV lies squarely in the
cosmologically forbidden mass range for stable active neutrinos $\nu$
\cite{11}.
The existence of such a massive active neutrino is also
disfavoured by the Super-Kamiokande data
\cite{fuk}.
However,
 as shown by Shi and Fuller
\cite{12}  for an initial lepton asymmetry of $\sim
10^{-3}$, a sterile neutrino $\nu_{s}$ of mass $m_{s} \sim 10$ keV
may be resonantly produced in the early universe
with near closure density, i.e., $\Omega \simeq 1$.
The resulting energy spectrum is not thermal
but rather cut off so that it
approximates a degenerate Fermi gas.
Sterile neutrinos in this mass range are also
constrained
by astrophysical bounds on
the radiative decay
$\nu_{s} \rightarrow \nu \gamma$ \cite{13}.
However, the allowed parameter space includes $m_{s} \simeq 15$ keV,
contributing $\Omega_{s} \simeq 0.3$
to the critical density,
as favoured by the BOOMERANG data
\cite{14}.

The statics of degenerate fermion stars is well understood, being the
Op\-pen\-hei\-mer-Volkoff equation in the relativistic case \cite{8}, or the
Lan\'{e}-Emden equation with polytropic index $n = 3/2$ in the
nonrelativistic limit \cite{15}. Alternatively,
one may understand these as the Thomas-Fermi theory applied
to self-gravitating fermion systems.
The extension of the Thomas-Fermi theory
to finite temperature \cite{16,23} has been used
to show that, at a certain
critical temperature, weakly interacting massive fermionic matter
undergoes a first-order gravitational phase transition from a diffuse
to a clustered state, i.e. a nearly degenerate fermion star.
However, such
studies do not  bear on the crucial dynamical question of
whether the fermion star can form through gravitational collapse of
density fluctuations in an orthodox cosmological setting.
Indeed,
since collisional damping is negligible, one would expect that only a
virialized cloud results \cite{11}.

$N$-body simulations of the collisionless Boltzmann or Vlasov equation
evidence a rather different picture: the collapse is followed by a
series of bounces, with matter expelled at each bounce, leaving behind
a condensed object \cite{18}.
By Liouville's theorem, the Vlasov
equation describes an incompressible fluid in phase-space so that it
respects a form of the exclusion principle.
Hence, these $N$-body
simulations are effectively fermion simulations.
What transpires is
that gravity, being attractive, self-organises the phase-space fluid
into a high-density/momentum core at the expense of other
low-density/momentum regions as seen in the evolution of the spherical
Vlasov equation \cite{19}.

Much the same behaviour is observed in the formation of mini-boson
stars through so-called gravitational cooling \cite{20}.
Such a boson
star is stable by balancing the
uncertainty and gravitational pressures.
A similar mechanism works in the presence of a quartic self-interaction
\cite{21}, which dominates over uncertainty pressure, resulting in an
equilibrium radius $R \gg 1/m$, where $m$ is the boson mass
\cite{22}.
Hence, we have a universal description of the physics underlying
the formation process: once the collapse proceeds far enough,
uncertainty, interaction or degeneracy pressure results in a bounce,
in which
the outgoing shock wave carries away the binding energy.
The virial counter argument mentioned
 above is bypassed, because the ejected matter invalidates
its assumption that there is no flow through the boundary.

The purpose of
this letter is to verify this picture for the formation of a
fermion star from a cold gravitationally unstable configuration.
 The dynamical
Thomas-Fermi theory, developed long ago by Bloch for the electron gas
\cite{23}, amounts to Euler's equations for irrotational flow,
subject to an equation of state $P = P(\rho)$.
The problem is that,
due to the presence of shocks
and instabilities,
a naive integration of the Bloch equations is
precluded
in the gravitational case.
  The usual remedy is to introduce some small numerical
viscosity. However, it seems imprudent to draw conclusions based on
introducing an {\em ad-hoc}
dissipation into what is fundamentally a dissipationless
system. Here we take another, literally conservative
approach.

We base our approach on the equivalence of a degenerate
non-interacting fermion star
with a certain type of self-interacting cold boson star.
We shall demonstrate this equivalence using scaling arguments
which are similar in spirit
to the analysis of self-interacting boson stars
by Colpi, Shapiro and Wassermann
\cite{22}.
Consider a complex scalar field $\psi$ with a
repulsive Lagrangean interaction term $U(|\psi|^2)$.
For simplicity, we use the Newtonian approximation,
but our analysis may be easily extended in a general relativistic
context.
In the Newtonian limit, a self-interacting boson star is governed by
the Gross-Pitaevskii-like equations
%.............
\begin{equation}
i \frac{\partial \psi}{\partial t}  = \left[ - \frac{\Delta}{2 m} +
\frac{1}{m} \frac{\d U}{\d |\psi|^2}
  + m \varphi \right] \, \psi,
 \label{dpsi}
\end{equation}
\begin{equation}
\Delta \varphi  =  4 \pi G   \rho,
 \label{delta}
\end{equation}
\begin{equation}
\rho=m^2|\psi|^2.
\label{rhopsi}
\end{equation}
The pressure tensor is given by
\begin{equation}
\label{eq102}
P_{ij}=
  \mathrm{Re}\,
\frac{\partial\psi}{\partial x_i}
\frac{\partial\psi^*}{\partial x_j}
+\delta_{ij}\left(|\psi|^2\frac{\d U}{\d |\psi|^2} -U
-\frac{\Delta|\psi|^2}{4}\right).
\end{equation}
Following Colpi et al. \cite{22}, we introduce
a dimensionless parameter $\Lambda$ and define
\begin{equation}
\label{eq104}
R_{*} \equiv \frac{\Lambda}{m} \equiv
\lambda\frac{m_{\mathrm{Pl}}}{m^2};
\;\;\;
M_{*}  =  R_{*} m^2_{\mathrm{Pl}}\,,
\end{equation}
as the length and mass scales, respectively.
Here $m_{\mathrm{Pl}} \equiv 1/\sqrt{G}$ denotes the Planck mass
and the arbitrary constant $\lambda$
will be fixed later using physical arguments.
%EQ 8
The substitution
\begin{equation}
\psi  =  \frac{m}{\sqrt{4\pi}\lambda} \Psi
\end{equation}
yields the coupled dimensionless equations
%EQ 9
\begin{equation}
\frac{i}{\Lambda} \; \frac{\partial \Psi}{\partial t}  =
\left[ - \frac{\Delta}{2 \Lambda^{2}} + \varphi +
V(|\Psi|^2)
 \right]  \Psi,
\label{eq009a}
\end{equation}
\begin{equation}
\label{eq009b}
\Delta \varphi  =   |\Psi|^2 ,
\end{equation}
where we have introduced the dimensionless potential
\begin{equation}
V(|\Psi|^2)=
\frac{4\pi\lambda^2}{m^4} \frac{\d U}{\d |\Psi|^2}.
\end{equation}
The  pressure tensor (\ref{eq102}), written in terms of
dimensionless variables, reads
\begin{equation}
\label{eq109}
P_{ij}=
 \frac{m^4}{4\pi\lambda^2}\frac{1}{\Lambda^2}
  \left(
   \mathrm{Re}\,
\frac{\partial\Psi}{\partial x_i}
\frac{\partial\Psi^*}{\partial x_j}
-\delta_{ij}\frac{\Delta|\Psi|^2}{4} \right)
+\delta_{ij}\left(|\Psi|^2\frac{\d U}{\d |\Psi|^2} -U\right).
\end{equation}
A static, spherically symmetric solution,
usually referred to as boson star, is obtained
by the ansatz
\begin{equation}
\Psi =\e^{-{\mathrm i}\epsilon R_* t}\Phi(r),
\label{ansatz}
\end{equation}
where $\Phi(r)$ is a real function.
It is clear from the definition (\ref{eq104})
that $\Lambda \gg 1$, as long as $m/m_{\mathrm{Pl}}\ll \lambda$
which is a reasonable assumption.
Hence, for sufficiently large $\Lambda$,
Eq. (\ref{eq009a}) with (\ref{ansatz}) degenerates to
\begin{equation}
\label{eq105}
\frac{\epsilon}{m} -\varphi -
V(\Phi^2)=0\, .
\end{equation}
This equation is exact in the limit
$\Lambda\rightarrow \infty$, which is just
the Thomas-Fermi limit \cite{24}.
In this limit
the derivative terms in Eq. (\ref{eq109}) vanish
so that the pressure tensor becomes diagonal,
with all the components equal to
$P\equiv P_{ii}$.
We obtain an effective equation of state given by
\begin{equation}
\label{eq106a}
\rho = \frac{m^4}{4\pi\lambda^2}\Phi^2,
\end{equation}
\begin{equation}
\label{eq106b}
P=
 \rho V -U.
\end{equation}
It may be easily shown that the last equation
combined with (\ref{eq105}) yields the equation
of hydrostatic
equilibrium
\begin{equation}
\label{eq107}
 \frac{dP}{\rho}=-d\varphi.
\end{equation}
For a particular  $U(|\psi|^2)$, the equation of
state in the form $P=P(\rho)$ is obtained by eliminating $\Phi^2$ from
(\ref{eq106a}) and
(\ref{eq106b}).
On the other hand,
if the  equation of state,
e.g., of the polytropic type,
is given,
we can determine the potential $U$ and its
derivative $V$
by integrating  (\ref{eq107}).
For a general polytropic equation of state
%EQ 5
\begin{equation}
P (\rho)  =  K  \rho^{1+1/n},
\end{equation}
Eq. (\ref{eq106b}) yields
%EQ 6
\begin{equation}
V=(n + 1)  K  \rho^{1/n} \; .
\label{eq6}
\end{equation}
This, together with Eq. (\ref{eq106a}), gives the potential in terms
of $\Phi^2$.
As we still have the freedom to choose
the parameter $\lambda$ conveniently,
we fix $\lambda$ so that
the potential $V$ in (\ref{eq009a}) takes a simple form
\begin{equation}
\label{eq108}
V=|\Psi|^{2/n} .
\end{equation}
The polytropic equation of state with $n=3/2$, together with
the equation of hydrostatic equilibrium (\ref{eq107}) and
Poisson's equation (\ref{delta}), describes a degenerate
fermion star.
Hence, we have demonstrated that a
  degenerate fermion star
is equivalent to a self-interacting boson star
in the limit $\Lambda\rightarrow \infty$.
Moreover, it may be shown numerically that,
even for moderate values of $\Lambda$ of the order of 10 to 20,
the static solutions
are almost degenerate and
are quite well
approximated by
the static solution for an infinite $\Lambda$.
This has been demonstrated
for a quartic self-interaction \cite{22}
(note that our $\Lambda$  corresponds
  to $\sqrt{\Lambda}$ in reference \cite{22}).

Next we proceed to  solving
Eqs. (\ref{eq009a}),
(\ref{eq009b})
and
(\ref{eq108})
numerically
for large, but finite value of
$\Lambda$.
For weakly interacting degenerate fermions with
the polytropic index  $n = 3/2$,
the length and mass scales are given by
\begin{equation}
R_{*}  =  \left( \frac{9 \pi^{2}}{32 g_{f}^{2}} \right)^{1/4} \;
\frac{m_{\mathrm{Pl}}}{m_{f}^{2}} = 0.2325
\left( \frac{\mbox{keV}}{m_{f}} \right)^{2}  \sqrt{\frac{2}
{g_{f}}} \; \mathrm{lyr} \; ,
\end{equation}
\begin{equation}
M_{*}   =  1.489 \left( \frac{\mathrm{keV}}{m_{f}} \right)^{2}
 \sqrt{ \frac{2}{g_{f}}} \times 10^{12} \; M_{\odot} \; ,
\end{equation}
where $g_{f}$ is the spin degeneracy and $m_f$ the
 mass of the fermion.
 As $M_*$ is of the order of the
 Oppenheimer-Volkoff limit,
the validity of the Newtonian
approximation in the static case requires
\begin{equation}
M=(4 \pi)^{-1}  \int  \d^{3}r \, | \Psi |^{2} \ll 1 .
\end{equation}

 By construction,
once the overall scale is specified,
a large but finite $\Lambda$ allows us to
simulate the fermionic problem as a bosonic one,
as long as their
Thomas-Fermi limits coincide,
while providing an explicitly energy conserving
way of controlling the shocks
and instabilities.
The basic regulating mechanism is the
kinetic part of (\ref{eq009a})
which penalises density spikes.
Of course,
$\Lambda$ must be so large that this term does not change
the static scaling relationship \cite{15}
\begin{equation}
M  R^{\left( \frac{3-n}{n-1} \right)}  =  C_{n},
\label{eq012}
\end{equation}
arising from the polytropic equation of state.
Our criterion is that
the ratio of kinetic and pressure contributions to the static
energy functional should be small.
In particular for
a Gaussian $\Psi
= \alpha\exp\left[ - (r/\beta)^{2}\right]$
we find
%EQ 13
\begin{equation}
\frac{1}{2\Lambda^2}
\frac{\int \d^3r |\nabla \Psi|^2}{
\int \d^3r |\Psi|^{2+2/n}}
=  \left( 1 + \frac{1}{n} \right)^{3/2}
\frac{3}{2  \Lambda^{2} \beta^{2}  \alpha^{2/n}} \; .
\end{equation}
For $n = 3/2$ this is
independent of the size of  $\beta$ for a given mass,
yielding the weak condition $M \gg 0.91 \Lambda^{-3}$.
When the mass violates this inequality
the self-interaction $V$ becomes irrelevant and
there is a crossover to mini-boson star behaviour
$M R=$ const.

\begin{figure}[h]
\centering
\epsfig{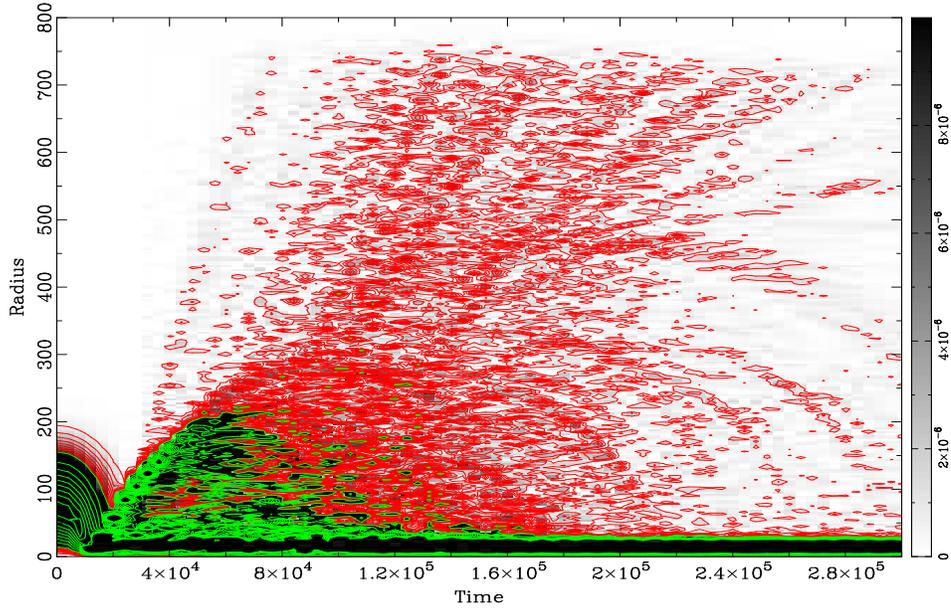}
\caption{
Combined
contour-density plot for the evolution of $|r \Psi |^{2}$
           from the initial configuration
           described in the text.
Green contour lines denote levels
from $10^{-5}$ to $10^{-4}$ while red lines denote
levels below $10^{-5}$.
%The lightest gray area correspond to a threshold value of
%$10^{-6}$  while the darkest area corresponds to levels
%at or exceeding $10^{-5}$.
Gravitational collapse is followed by ejection of
excess matter leaving a fermion star at the center.}
\label{fig1}
\end{figure}

In Fig.\ \ref{fig1} we display the evolution of $|r \Psi |^{2}$ for
the spherical collapse of a mass $M=0.008$, initially in the form of
the above
Gaussian with $\beta = 100 = \Lambda$.
A fairly coarse sampling is
used to make this contour plot, as this allows one to see the basic
features of matter ejection without the complicated fine detail hiding
it. Eqs. (\ref{eq009a}) and
 (\ref{eq009b})
are first written in terms of $\eta = r
\Psi$, and then solved assuming spherical symmetry using the method of
lines \cite{25} for $r = 0$ to 1000. The region is divided into 4000
intervals,
and a fourth order polynomial is fitted in each interval, with
the function continuous up to its second derivatives across each
interval.
The time step is adjusted dynamically to achieve the desired
accuracy of $10^{-6}$. The boundary conditions at $r = 0$ and $r =
1000$ are reflective. To prevent matter from being artificially
reflected by the boundary at $r = 1000$, we have
introduced an
$r$-dependent imaginary part to the potential,
or `sponge' \cite{21},
from $r =$ 700  to 1000, which removes the ejected fermionic matter.
The mass in the region remains
constant to within the numerical integration
error until $t = 5 \times 10^{4}$, which is a quarter of the
simulation period,
when the first ejected matter reaches the
`sponge'.
The expected features of bounce and ejection, leaving a
condensed core, are evident.
We  obtain identical results using
the Crank-Nicholson method.

\begin{figure}
\centering
\epsfig{file=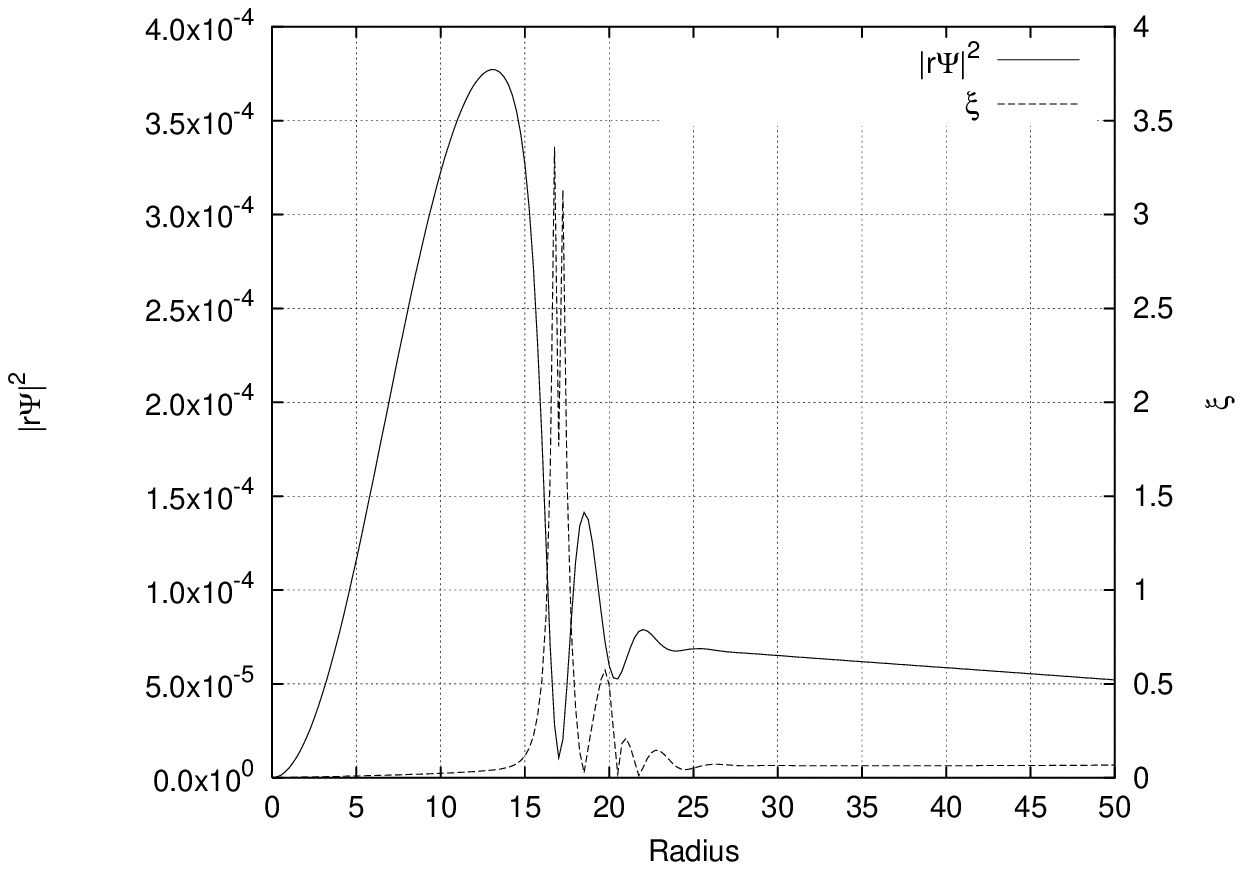,height=9.0cm}
\caption{$| r \Psi |^{2}$ (solid line) and $\xi$ (dashed line)
           versus $r$ at $t = 1.22 \times 10^{4}$.}
\label{fig2}
\end{figure}
\begin{figure}
\centering
\epsfig{file=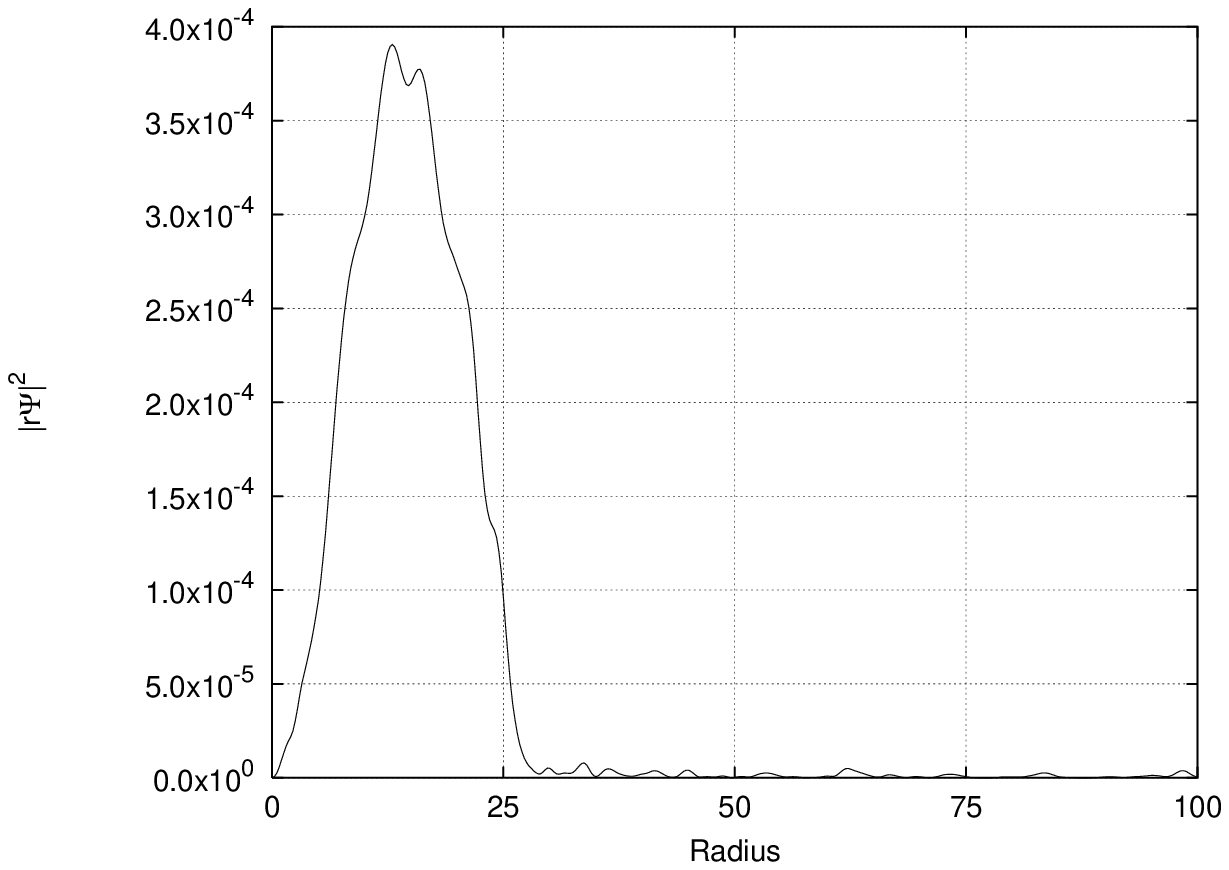,height=9.0cm}
\caption{ $| r \Psi |^{2}$  versus $r$ on the time slice
$t=2\times 10^5$
          in Figure 1.}
\label{fig4}
\end{figure}

A pertinent question is whether the Thomas-Fermi
approximation
is satisfied during the ejection process.
Similar to WKB,
the Thomas-Fermi limit
is obtained when the scale of variation of density
is small compared to that of the potential $V$, i.e.,
\begin{equation}
\xi^2 \equiv
\frac{ |\nabla\rho|^2}{
\rho^2\, 8\, \d U/\d |\psi|^2}=
\frac{|\nabla |\Psi||^2}{
2\Lambda |\Psi|^{2+2/n}}
\ll 1.
\end{equation}
In Fig.\ \ref{fig2} we
exhibit $|r \Psi|^{2}$ and $\xi$ versus $r$ at time $1.22
\times 10^{4}$, where ejecta first develop.
Evidently, the Thomas-Fermi
condition, is well satisfied, except near sharp density minima.
This is expected since in a pure Thomas-Fermi system evolving under
Bloch's equations, the sharp density gradient would evolve into a shock,
which violates the Thomas-Fermi condition.

In Fig. \ref{fig4} we show $|r \Psi|^{2}$ on the time slice
$t=2\times 10^5$ of
our $\Lambda=100$ simulation.  The core mass $M = 0.0057$ and radius
$R = 28$ are roughly commensurate with (\ref{eq012}) and \cite{15}
$C_{3/2} = 132.3843$. Of course, at this point there is still evolution
and oscillation of the core, similar to the mini-boson star case
\cite{20}, with $R$ varying between about 25 and 31. In general one can
expect a persistent seismology with a period $T$ of the order of the
radius divided by the average speed of sound, $T \sim 51/M$, which
agrees with the simulations. The seismology includes
higher-order modes as well.

In summary, using a bosonic representation of the
dynamical Thomas-Fermi
theory for a self-gravitating gas, we have shown that nonrelativistic,
degenerate and weakly interacting
fermionic matter will form supermassive
fermion stars through gravitational collapse accompanied by ejection.
For a fermion mass of
$m_{f} \simeq 15$ keV,
and a final mass $M\simeq 2.6\times 10^6 M_{\odot}$
such a superstar is consistent
with the observations of the compact dark object at
the center of our galaxy.
A similar demonstration for the formation
of such a star near the Oppenheimer-Volkoff limit, and
the question of cosmology with
degenerate dark matter, requires a general
relativistic extension which is under
development and will be reported
elsewhere.

%\newpage
%________________________
%{\bf Acknowledgement}

\begin{ack}
The authors wish to thank Duncan Elliott ($\dag$)
%\footnote{Deceased 20 July 2000}
for many useful discussions regarding the simulations.
This
research is in part supported by the Foundation of Fundamental
Research (FFR) grant number PHY99-01241 and the Research Committee of
the University of Cape Town.  The work of N.B. is supported in part by
the Ministry of Science and Technology of the Republic of Croatia
under Contract No. 00980102.
\end{ack}

\end{document}